# Venus Life Finder Missions Motivation and Summary

Sara Seager [1,2,3,*], Janusz J. Petkowski [1], Christopher E. Carr [4], David H. Grinspoon [5], Bethany L. Ehlmann [6], Sarag J. Saikia [7], Rachana Agrawal [8], Weston P. Buchanan [8], Monika U. Weber [9], Richard French [10], Pete Klupar [11], Simon P. Worden [11], Darrel Baumgardner [12,13] and on behalf of the Venus Life Finder Mission Team [†]

[1] Department of Earth, Atmospheric and Planetary Sciences, Massachusetts Institute of Technology, 77 Massachusetts Avenue, Cambridge, MA 02139, USA; jjpetkow@mit.edu
[2] Department of Physics, Massachusetts Institute of Technology, 77 Massachusetts Avenue, Cambridge, MA 02139, USA
[3] Department of Aeronautics and Astronautics, Massachusetts Institute of Technology, 77 Massachusetts Avenue, Cambridge, MA 02139, USA
[4] School of Aerospace Engineering and School of Earth and Atmospheric Sciences, Georgia Institute of Technology, Atlanta, GA 30332, USA; cecarr@gatech.edu
[5] Planetary Science Institute, 1700 East Fort Lowell, Suite 106, Tucson, AZ 85719-2395, USA; grinspoon@psi.edu
[6] Division of Geological and Planetary Sciences, California Institute of Technology, Pasadena, CA 91125, USA; ehlmann@caltech.edu
[7] Spacefaring Technologies Pvt. Ltd., 15F, No. 14, Bhattrahalli Old Madras Road, KR Puram, Bangalore, 560049, Karnataka, India; saragjs@gmail.com
[8] School of Aeronautics and Astronautics, Purdue University, 701 W. Stadium Ave., West Lafayette, IN 47907, USA; rachna.agrawal.04@gmail.com (R.A.); buchanaw@purdue.edu (W.P.B.)
[9] Fluid-Screen, Inc., 100 Cummings Center, Suite 243-C, Beverly, MA 01915, USA; monika.weber@fluid-screen.com
[10] Rocket Lab, 3881 McGowen Street, Long Beach, CA 90808, USA; r.french@rocketlabusa.com
[11] Breakthrough Prize Foundation, NASA Research Park, Building 18, P.O. Box 1, Moffett Field, CA, 94035-0001, USA; klupar@breakthrough-initiatives.org (P.K.); pete@breakthroughprize.org (S.P.W.)
[12] Droplet Measurement Technologies, LLC, 2400 Trade Centre Ave, Longmont, CO, 80503, USA; darrel.baumgardner@gmail.com
[13] Cloud Measurement Solutions, LLC, 415 Kit Carson Rd, Unit 7, Taos, NM, 87571, USA
* Correspondence: seager@mit.edu
† Venus Life Finder Mission Team. All memberships are listed in Acknowledgments.

**Abstract:** Finding evidence of extraterrestrial life would be one of the most profound scientific discoveries ever made, advancing humanity into a new epoch of cosmic awareness. The Venus Life Finder (VLF) missions feature a series of three direct atmospheric probes designed to assess the habitability of the Venusian clouds and search for signs of life and life itself. The VLF missions are an astrobiology-focused set of missions, and the first two out of three can be launched quickly and at a relatively low cost. The mission concepts come out of an 18-month study by an MIT-led worldwide consortium.

## 1. The Venus Opportunity

The concept of life in the Venus clouds is not new, having been around for over half a century (e.g., [1–7]). What is new is the opportunity to search for habitable conditions or signs of life directly in the Venus atmosphere with scientific instrumentation that is both significantly more technologically advanced and greatly miniaturized since the last direct in situ probes to Venus' atmosphere in the 1980s.

The idea of life in the Venus atmosphere is highly controversial. The Venus cloud environment is very harsh for life of any kind. The clouds are composed of concentrated sulfuric acid ($H_2SO_4$) particles that are orders of magnitude more acidic than the most acidic environments where life is found on Earth. The cloud layers are about 50 to 100 times drier than the Atacama Desert, one of the driest places on Earth and far drier than the limits of life as we know it [5,8]. Whether or not Venus had ancient water oceans where life as we know it could have originated and later migrated to and evolved in the clouds is under debate [9–11].

Nonetheless, Venus is a compelling planet to search for signs of life because of: the habitable temperatures in the cloud layers; the many atmospheric chemical anomalies suggestive of unknown chemistry (Figure 1 and [12]); and new laboratory experiments on organic chemistry in sulfuric acid [13].

The VLF series of missions (described in [14] and in companion papers in the same issue [15–17]) are directly formulated to assess the habitability of Venusian clouds and to search for signs of life and life itself (Table 1). The VLF missions are a focused set of missions for which the first two can be launched quickly and executed with relatively low cost. While NASA and ESA have recently selected missions to visit Venus at the end of the 2020s (VERITAS [18], DAVINCI [19], and EnVision [20]), these missions are for general studies about the planet's properties and do not address the habitability and astrobiology questions targeted by the VLF missions.

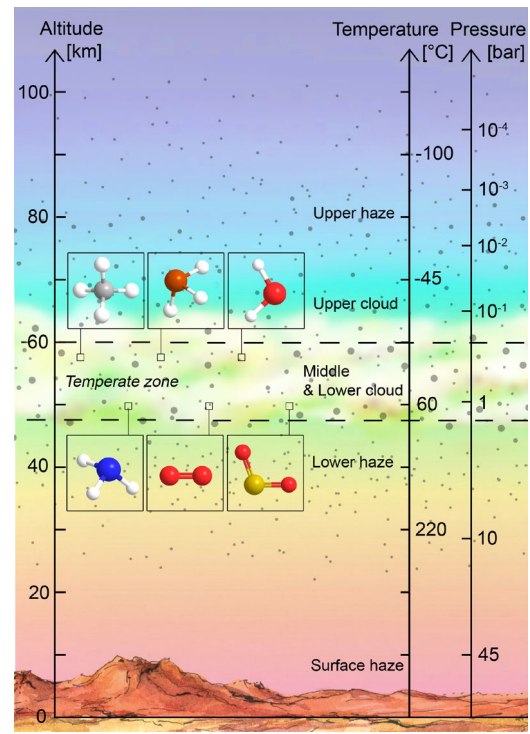

**Figure 1.** Schematic of Venus' atmosphere. The cloud cover on Venus is permanent, continuous, and vertically extensive. The middle and lower cloud layers have temperatures suitable for life. Selected molecules relevant to habitability or life and with unexplained presence or vertical abundances are shown. Atom colors are as follows: H = white; C = grey; P = orange; O = red; N = blue; S = yellow. Figure adapted from [5].

**Table 1.** Summary of VLF mission concepts, science goals, objectives, and instruments. Instruments: TLS = Tunable Laser Spectrometer, MoOSA = Molybdenum Oxide Sensor Array, TOPS = Tartu Observatory pH Sensor, MEMS = Microelectromechanical Systems, MEMS-A = MEMS Aerosol Elemental Analyzer, MEMS-G = MEMS Gas Molecule Analyzer, AFN = Autofluorescing Nephelometer, LDMS = Laser Desorption Mass Spectrometer, FSC = Fluid-Screen Concentrator, AMP = Autofluorescing Microscope System. Missions: RL = Rocket Lab Mission, HAB = VLF Venus Habitability Mission, VAIHL = VLF Venus Airborne Investigation of Habitability and Life Mission. See also [14] and companion VLF papers in the same Special Issue [15–17].

| | Goals | Science Objectives | Instruments | Mission Concept |
|---|---|---|---|---|
| Habitability | 1. Measure Habitability Indicators | 1.1 Determine the amount of water in the cloud layers | TLS and Conductivity Sensor | HAB, VAIHL |
| | | 1.2 Determine the pH of single cloud particles | MoOSA and TOPS acidity sensors | HAB, VAIHL |
| | | 1.3 Determine and identify metals and other nonvolatile elements in the cloud particles | MEMS-A, LDMS | HAB, VAIHL |
| | | 1.4 Measure the temperature, pressure, and wind speed as a function of altitude | Temperature and pressure sensor, anemometer | HAB, VAIHL |
| Biosignatures | 2. Search for Evidence of Life in the Venusian Clouds | 2.1 Search for signs of life via gas detection | TLS and MEMS-G | HAB, VAIHL |
| | | 2.2 Detect organic material within the cloud particles | AFN, FSC with AMP | RL, HAB, VAIHL |
| | | 2.3 Identify organic material within the cloud particles | AFN, MEMS-A, LDMS | HAB, VAIHL |
| | | 2.4 Detect and characterize morphological indicators of life | FSC with AMP | VAIHL |
| Sample Return | 3. Characterize Cloud Particles in Preparation for Sample Return | 3.1 Determine if the cloud particles are liquid or solid | AFN | RL, HAB, VAIHL |
| | | 3.2 Determine if the cloud particles are homogeneous | AFN, acidity sensors, MEMS-A, LDMS | RL, HAB, VAIHL |

Remarkably, it has been nearly 40 years since the last Venus in situ measurements. The Soviet Union's missions including landers and two balloons flew from the late 1960s through the early 1980s and the US Pioneer Venus probes flew in 1978. The entire scientific field of Astrobiology has sprung up in the interim. The VLF missions to Venus will take advantage of an opportunity for focused and high-reward science, which stands to possibly answer one of the greatest scientific mysteries of all, and in the process pioneer a new model of public–private partnership in space exploration.

## 2. New Findings for Life's Survival on Venus

The VLF mission team set out to take a fresh look at whether the harsh atmospheric conditions at Venus could be suitable for single-celled microbial life. Of particular interest are the scarcity of water in the atmosphere and the extreme acidity of the sulfuric acid cloud droplets. We designed and performed chemistry and biology experiments to guide mission science objectives related to habitability and the search for life.

First, we set out to show that the highly acidic droplets can support organic chemistry and are not necessarily sterile "dead" zones from an organic chemistry perspective. The team seeded test tubes of concentrated sulfuric acid with simple organic molecules. The result was an intriguing rich and complex set of organic molecules. What this shows is that sulfuric acid droplets can support a variety of complex chemicals that can in principle lead to the formation of the building blocks of exotic biochemistry based on concentrated sulfuric acid [13]. Such results are very intriguing because all life needs complex organic chemistry. This result is also seen in the petrochemical industry where "red oil" is a waste byproduct of fuel production.

Second, we aimed to find life biochemical materials that could survive in concentrated sulfuric acid. Although sulfuric acid is known to be destructive to many biochemicals and engineering materials, there are some that are resistant. We discovered a set of lipids that can not only survive in sulfuric acid (tested at concentrations of 70% or lower) but can self-assemble to form vesicle-like structures (described in [14] Appendix A). More work is needed to determine the membrane permeability and other chemical properties. A membrane is critical to protect life's biochemicals from the outside environment.

Third, we propose that locally and biologically produced ammonia ($NH_3$) can neutralize the Venus sulfuric acid cloud droplets such that a subset of the cloud particles may be brought to an acidity level tolerable by acidophiles on Earth [21]. Concentrated sulfuric acid is billions of times more acidic than the most acidic environment that is inhabited by life on Earth. Yet with $NH_3$ neutralization, the Venus cloud particles will be brought to about pH = 0 or 1. This theory is motivated by the suggested presence of $NH_3$ by both Venera 8 [22] and the Pioneer Venus probe measurements [23] and leads to a cascade of reactions that help resolve long-standing Venus atmosphere anomalies [21]. The anomalies include: the large particles, called "Mode 3", in the lower cloud layers that appear to be non-spherical and therefore cannot be pure concentrated sulfuric acid which is a liquid that would form spherical droplets; the gas $O_2$ found at significant levels (10s of ppm); depletion of $SO_2$ and $H_2O$ in the cloud layers; and the possible presence of $NH_3$. On Earth, co-occurring $NH_3$ and $O_2$ are only associated with life as both gases require continuous and efficient production in significant amounts to offset their reactivity in the atmosphere (see e.g., [24,25]). This new theory that life-produced $NH_3$ sets off a chain of reactions that explains Venus's atmosphere anomalies may be countered by the idea of unknown chemical processes at play for each individual anomaly. One such possibility is that the $SO_2$ depletion alone could be explained by cloud droplet chemistry and acid neutralization by minerals lofted from the surface if such minerals reach the clouds in sufficient abundance [26]. New chemistry itself is of significant interest to planetary science.

Based on the above three points, our mission priorities include the search for organic compounds in the cloud particles, the measurement of the acidity of the cloud particles, and a sample return to search for cell-like structures and complex organic biopolymers. In addition, the missions will search for biosignature gases and other indicators of habitability or life.

## 3. A Small Venus Atmosphere Probe for a Targeted 2023 Launch

We partnered with Rocket Lab such that the VLF team will provide the science payload and science team for a mission to Venus with a target launch date of May 2023, and a backup launch date of January 2025. Rocket Lab is providing the Electron launch vehicle, the cruise phase Photon spacecraft, and the atmospheric probe entry vehicle. The entry vehicle flying aboard Rocket Lab's Photon spacecraft on a direct entry to Venus has room for up to 1 kg of scientific instrumentation for the short-duration (three-minute) descent through the cloud layers.

The primary science goal is to search for organic compounds in Mode 3 or other cloud particles. A discovery of organic compounds would show that complex molecules suitable for life could exist. If life is present, it is most likely microbial-type life residing inside the cloud particles. This first probe mission would not be able to identify such life but could indicate the presence of organic molecules. Organic compounds with delocalized electrons in ring structures, when subjected to UV light, yield stronger fluorescent signals compared to other molecules.

The science instrument is the Autofluorescing Nephelometer (AFN). The AFN will shine a UV laser through a probe window to induce autofluorescence in any organic material inside cloud particles. If there is no autofluorescence detected, the AFN will still return useful science by the secondary science goal. A measurement of the intensity of the laser light and polarization backscattered off of the particles will be used to constrain the composition and shape of the particles. If we can confirm past measurements that indicate some cloud particles are non-spherical, i.e., not liquid, it affirms the possibility that some particles are not made of liquid concentrated sulfuric acid and therefore could be more habitable to life as we know it than previously thought. The non-spherical particles would represent a currently unknown composition, including the possibility of biologically produced ammonium salt slurries [21,27].

The AFN is being built by Droplet Measurement Technologies (DMT) and has a high heritage via DMT's commercial products that fly on the outside of aircraft. The backscatter cloud probe (BCP) [28] has over 60,000 h of data acquired on a global basis [29]. The BCP was subsequently upgraded to include the detection of polarization, and the AFN takes this further by the addition of a fluorescence detection channel. The laboratory tests on AFN detection of fluorescence of organic compounds, including potential contaminants and mineral "false positives", are currently underway and are aimed to guide the data analysis and interpretation. Contamination from the probe heat shield (carbon phenolic) will be avoided by probing particles outside of the ablation airflow. However, extensive laboratory testing of many different types of particle compositions related to the expected ablation products will be carried out.

The Venus atmospheric entry probe will spend approximately three minutes in the cloud layers, after being deployed from the Photon spacecraft, and will make continuous measurements with the AFN, as well as obtain the pressure, altitude, and inferred temperature profile of the atmosphere. The choice of the AFN delivers new science and is unique as compared to the recently selected NASA and ESA missions because none of these missions intentionally include in situ studies of Venus cloud particles.

**4. A Balloon Mission to Establish Habitability and Search for Signs of Life**

Our VLF Venus Habitability Mission concept is a 4 m diameter fixed-altitude balloon mission to the Venus cloud layers that will use a tailored set of small instruments to search for habitability and signs of life (Figure 2). The mission philosophy is to develop a near-term implementable mission.

The fixed altitude balloon would operate for one to two weeks in the middle to lower cloud region, at about 48 to 52 km altitude where the temperature is habitable (between ~87 °C and ~37 °C respectively [30]). The operational altitude of choice of 52 km is based on temperature, the presence of the population of Mode 3 particles identified by the Pioneer Venus probe [31–33], and the anomalous gas abundances (Section 2) [12]. Passive altitude variation due to updrafts, downdrafts, and thermal effects as well as horizontal variation due to winds, will enable the measurement of spatial distributions of gases. The mission will deploy four mini probes from the balloon gondola to sample lower cloud layers.

The mission will: support or refute evidence for signs of life in the Venus cloud layers; ascertain the habitability of the Venusian clouds or lack thereof; and inform an atmospheric sample return (by determining cloud particle phase and homogeneity and balloon technology demonstration). The science payload has a mix of mature and novel instruments chosen to achieve the science goals, summarized as follows.

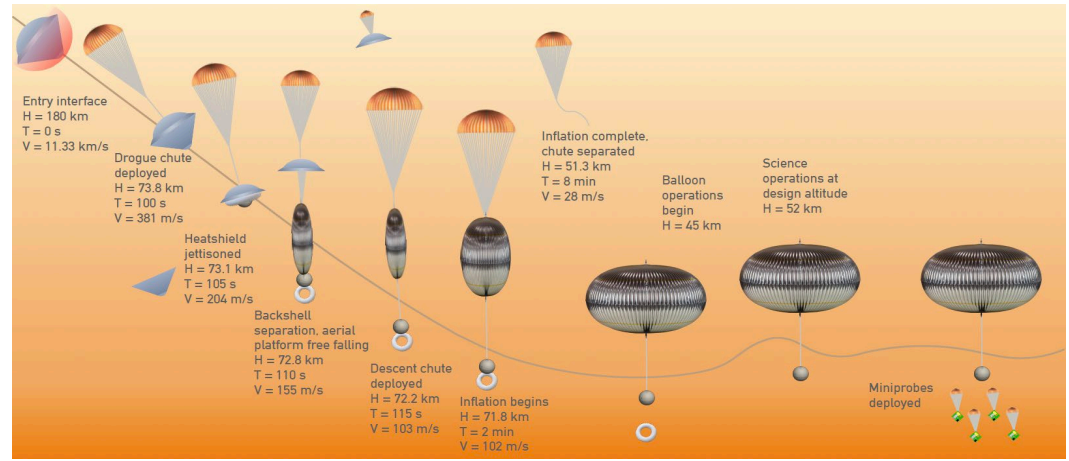

**Figure 2.** Balloon mission entry, descent, and deployment concept of operations. The 52 km fixed-altitude balloon mission is planned to be operational for one week and communicate both through an orbiter as well as direct communication to Earth. Mini probes will be deployed to sample the atmosphere below 52 km.

A four-channel mini Tunable Laser Spectrometer (TLS) will measure the abundances of key biosignature gases: $O_2$, $NH_3$, and $PH_3$, and the habitability indicator $H_2O$.

An Autofluorescence Nephelometer (AFN) with UV excitation capability will measure backscattered polarized radiation and any induced fluorescence. The AFN is a power-hungry instrument, with a peak operating power of 40 W, but it is small (~100 cm$^3$) and lightweight (~800 g) [16]. The AFN will be enhanced in capability compared to the Rocket Lab Mission AFN, by way of additional excitation lasers and a broader wavelength range detector. The AFN will also utilize an inlet to enable operation during night or day, whereas during the initial probe mission, the need to measure particles external to the probe requires a night entry.

A microelectromechanical systems (MEMS) device to detect non-volatile elements including metals, which are needed for all life as we know it to catalyze metabolic reactions [34,35]. The MEMS device will be tailored to elements of interest.

A single particle acidity sensor will investigate the hypothesis that cloud particles may have acidities of pH = 1. A measurement of pH = 1 would be a major discovery because pH = 1 is consistent with the environment for acidophiles on Earth whereas the acidity of concentrated sulfuric acid is orders of magnitude lower and destructive to all life as we know it. As part of this study we motivated the development of two independent single-particle pH sensors (e.g., based on the modified molybdenum oxide senor [36], see also [17] for an overview of the instruments considered for the VLF Venus Habitability Mission).

A weather instrument suite will measure temperature–pressure profiles and wind speed. While not directly astrobiological in nature, these are worth measuring in their own right. Transient planet gravity waves are encoded in the temperature–pressure profiles (e.g., [37]), and measuring them helps substantiate the concept of transporting materials, including those which might support the cycling of organic carbon and possibly life, up from lower atmosphere layers.

Four mini probes will be deployed from the balloon gondola to sample atmosphere layers below 52 km. Two mini probes would contain a pH sensor and two mini probes would contain a MEMS gas detector. Each would also contain the weather instrument suite.

The total balloon mass, i.e., the balloon envelope (15 kg) together with the gondola carrying the science payload and mini probes (30 kg), would be about 45 kg.

During the study, we considered a number of trades, including a variety of scientific instruments and two balloon categories. A fixed altitude balloon is simpler and less costly than a variable altitude balloon (up to a factor of five). Many of the desired instruments do not fit with the small balloon payload mass and need more development time to mature to flight readiness. For example, a laser desorption mass spectrometer is a powerful tool for the non-pyrolyzed identification of a wide variety of compounds [38,39]. A Fluid-Screen particle concentrator can capture enough particles for a microscope focal plane [40]. A liquid collector to feed a mass spectrometer and a microscope have to be developed and so far there is no suitable microscope for a balloon-borne platform that can reach down to the desired 0.2 µm minimum cell size for life.

As an alternative to a mission consisting of a balloon and mini probes, future work will develop a mission concept with two large identical probes with parachutes that might spend up to an hour descending through the cloud layers with the same instrument suite described above.

**5. A Venus Atmosphere Sample Return**

Ultimately a Venus atmosphere sample return is needed to robustly answer the compelling question, "Is there life in Venus' clouds?" The VLF Atmosphere Sample Return Mission aims to return up to one liter of gas and up to a few grams of cloud particles (Figure 3). The target cloud altitude region for sample collection will be informed by the results of both the Rocket Lab entry probe mission (Section 3) as well as the VLF Venus Habitability Mission (Section 4). A sample return without a doubt is the most robust way to search for signs of life or life itself in Venus' atmosphere. Earth-based laboratories include a wider variety of sophisticated tools with higher sensitivity than space-based instrumentation, and the presence of human investigators allows for a vastly wider range of potential experiments.

Prior to a sample return, we need to invest in sample capture and storage technologies. We need to understand whether or not a subset of the Venus cloud particles are solids vs. liquids because this informs the sample capture methods. We also need to investigate the homogeneity of cloud material in order to determine the viability of storing particles en masse. (Storing particles en masse, instead of individually, might change the chemical composition of the sample and obfuscate the true chemical environment of Venus' cloud particles.)

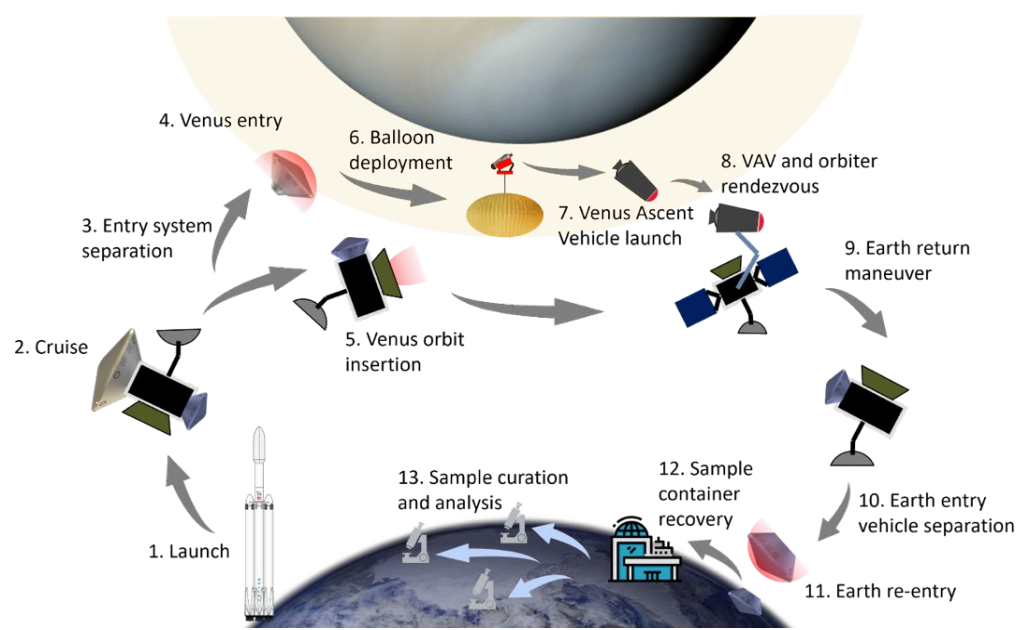

**Figure 3.** Schematic mission architecture for atmospheric sample return from Venus. VAV stands for Venus ascent vehicle.

A cost-effective balloon-borne mission to Venus will be able to establish the limits of habitability of the Venusian atmosphere and search for signs of life, as well as test technology needed for sample return. We have several international partners interested in joining our balloon mission team.

We recommend investment in technology for a sample return mission, specifically gas and liquid capture and storage technologies.

## 6. Summary

Collectively, our suite of three Venus astrobiology missions offers a focused high-impact route to seeking life beyond Earth, possibly enabling a truly historic first discovery. The near-term Rocket Lab entry probe mission offers the potential to uncover a smoking gun with regard to Venus cloud habitability at low cost while enabling instrumentation validation and team building that would support a more ambitious balloon mission. Both of these missions would precede and complement planned NASA and ESA missions while laying the groundwork for a Venus atmospheric sample return mission. Even if no life is present, current evidence suggests that we have much to learn about this extremely alien world.

**Author Contributions:** Conceptualization, S.S., J.J.P., C.E.C., D.H.G., B.L.E., S.J.S., R.A., W.P.B., M.U.W., R.F., P.K., S.P.W., D.B.; writing—original draft preparation, S.S., J.J.P.; writing—review and editing, S.S., J.J.P., C.E.C., D.H.G., B.L.E., S.J.S., R.A., W.P.B., M.U.W., R.F., P.K., S.P.W., D.B.; All authors have read and agreed to the published version of the manuscript."

**Funding:** This research was partially funded by Breakthrough Initiatives, the Change Happens Foundation, and the Massachusetts Institute of Technology.

**Institutional Review Board Statement:** Not applicable.

**Informed Consent Statement:** Not applicable.

**Data Availability Statement:** Not applicable.

**Acknowledgments:** We thank the extended Venus Life Finder Mission team for useful discussions. List of the individuals involved as the VLF extended Venus Life Finder Mission team can be found here: https://venuscloudlife.com/, accessed on 14th of May 2022.

**Conflicts of Interest:** The authors declare no conflicts of interest.